# Computer-Generated Photorealistic Hair

## Alice J. Lin

Department of Computer Science, University of Kentucky, Lexington, KY 40506, USA
ajlin0@cs.uky.edu

## Abstract

This paper presents an efficient method for generating and rendering photorealistic hair in two dimensional pictures. The method consists of three major steps. Simulating an artist drawing is used to design the rough hair shape. A convolution based filter is then used to generate photorealistic hair patches. A refine procedure is finally used to blend the boundaries of the patches with surrounding areas. This method can be used to create all types of photorealistic human hair (head hair, facial hair and body hair). It is also suitable for fur and grass generation. Applications of this method include: hairstyle designing/editing, damaged hair image restoration, human hair animation, virtual makeover of a human, and landscape creation.

## 1 Introduction

Because of the ubiquity of hair in everyday life, the human hair rendering has been an active area for over a decade. It is one of the most unsatisfactory aspects of rendered human images to date. Although there are some methods, which could render 3D hair, these rendering have always been hard. The hair they generated is not realistic appearance of human hair. And also, it is only for virtual human. This paper takes a different approach. It essentially uses image-processing techniques. This method has its own proprietary advantages. First, it greatly reduces the tasks of geometric modeling and of specifying hair color, shadow, specular highlights and varying degrees of transparency. It allows much more complicated models (such as curly hair and facial hair) to be generated and rendered. Second, it provides the flexibility of using any type of photograph, computer-generated image, or painting as input. Finally, it offers the ability that can much more easily modify the hairstyle, hair shape, hair color and thickness.

The past works of 2D hairstyle generation and design/change used the principle of the "cut and paste" (cutting the hair from one picture and pasting it to the other) approach with fixed hair models and fixed viewpoint pictures from the database. The results were far from being natural-looking and photorealistic. In this paper an integrated set of methods for generating photorealistic hair is presented. The approach allows us to create a variety of hairs. The paper focuses on human (head and facial) hair among the various types of hairs. The aim is to generate photorealistic hair on real human pictures by directly growing hair on the human head or face. The hair generated will naturally combine with the scene existing in the picture.

The paper is organized as follows. Section 2 introduces three procedures, and describes how they can be used to implement it. Section 3 discusses some of the results. Section 4 concludes and presents future work.

## 2 Generating photorealistic hair

To render photorealistic hair, we will consider these aspects: large number of hair, detailed individual hair and complex interaction of light and shadow among the hairs. The image of hair, in spite of the structural complexity and consisting of regions of hair color and shadow, shows a definite pattern and texture in its aggregate form. Improper rendering of this delicate detail can result in aliasing that causes the hair to look artificial. In this section, the basic idea of the technique is highlighted, which can easily handle these complex components for generating photorealistic hair. Figure 1 illustrates the framework of the method.

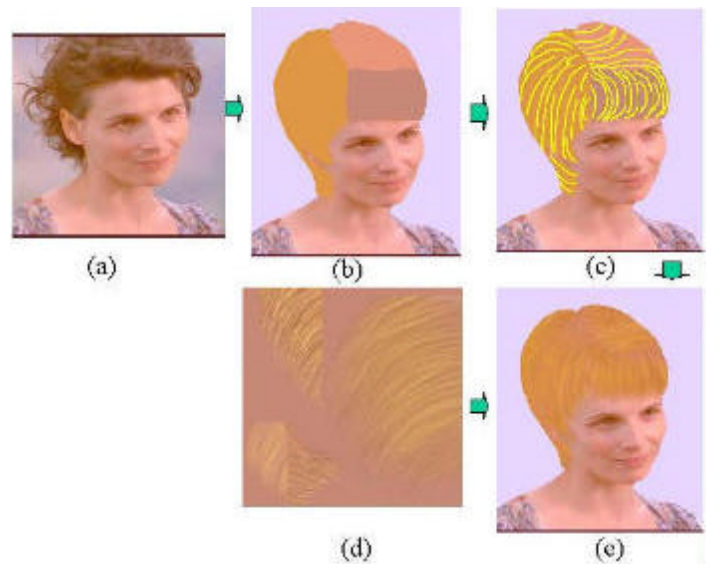

Figure 1: An overview of photorealistic hair generation. (a) Original picture. (b) Rough hair shape. (c) Stylized hair. (d) Hair patches generated. (e) New hair style.

The first step to create a new hairstyle is to design rough hair shape on original picture. After that, according to surface detail from the scalp, it is divided into several patches. Then the artist drawing is simulated to generate a cluster of hair, a strand of hair and individual hairs to sketch stylized hairs. On stylized hairs, filter-procedure is applied, which will turn the artificial-looking hair into realistic hair. Finally, refine-procedure is used to refine the boundaries of patches with the surrounding area.

### 2.1 Drawing - procedure

Simulating artist drawing achieves the basic hairstyles. The technique of pen-and-ink-style line drawing [1] [2] proposed by Salisbury et al is applied to drawing procedure. It can control the hair density, spread, color, width and length, reflecting light and casting shadow within many individual hairs. Hair strands and individual hairs can range from short straight hairs to long curly hairs of practically any design. Figure 2 is the example for drawing-procedure.

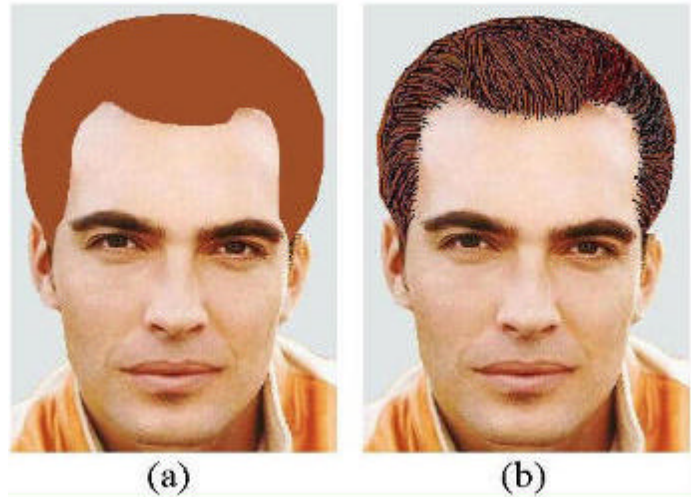

Figure 2: Drawing - procedure (a) Rough hair shape. (b) Drawn hair.

## 2.2 Filter-procedure

After drawing the basic hairstyle, the hair is non-photorealistic. The filter-procedure will turn the artificial-looking hair into realistic-looking hair. A convolution is carried out directly on an image by moving the convolution kernel so that it is centered on each pixel of the image in turn, then multiplying the corresponding elements and summing the products.

A convolution operation is used to blend well with the individual hair among other hairs, scalp and other things. Kernel specifies how a source pixel and its neighbors are combined. The center of the kernel represents the source pixel and the other elements correspond to the neighboring source pixels. The destination color is calculated by multiplying each pixel color by its corresponding kernel coefficient and adding the results together.

The convolution kernel is the key of generating realistic hair. The color of each destination pixel is determined by combining the colors of the corresponding source pixel and its neighbors. To preserve the brightness of the image, all elements of the kernel must add up to one. If they add up to more than one, the destination image will be brighter than the original. If the sum of the kernel coefficient is less than one, the destination image will be darker. This is because the colors of all the pixels used by the kernel are combined to form a single destination pixel color.

The coefficient values in the kernel are all less than or equal to one. It will not add energy (light, pixel levels) to the destination image. The coefficients and size of kernel control the hair's smoothness, density, and color intensity, and also affect curly direction, curly degree, etc. Figure 3 shows a real human scalp with stylized hair, applying the filter with 19*19 kernel and 31*31 kernel. In the Figure 3, (b) and (c), the kernel coefficients are assigned. White squares are for zero. Red squares are for positive values. (d) and (e) are the results of using different sizes of kernels.

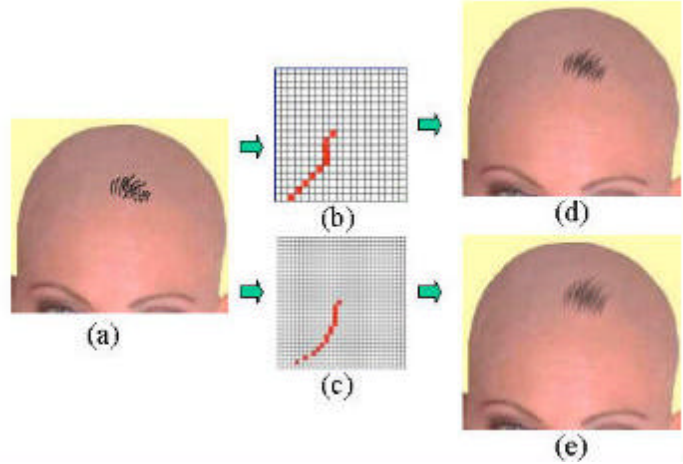

Figure 3: Filtering an image with different sizes and coefficients of matrixes. (a) A real human scalp with stylized hair. (b) 19 * 19 matrix used to generate the rough and short real hair. (c) 31* 31 matrix used to generate smooth and longer real hair. (d), (e) the results of filtering.

## 2.3 Refine--procedure

After the filter procedure the pixels of the patch are changed. If the pixel intensities of the patch have significant change, you wish to smooth out discontinuities between patches and the surrounding areas. The pixel intensity interpolation will be used to refine the patch boundary color with the surrounding areas. The equation for the pixel intensity interpolation is:

$$Ip = \frac{y - y2}{y1 - y2} \times Ip1 + \frac{y1 - y}{y1 - y2} \times Ip2 \qquad (1)$$

When the two points of p1 and p2 have same y value, it uses the equation (2).

$$Ip = \frac{x2 - x}{x2 - x1} \times Ip1 + \frac{x - x1}{x2 - x1} \times Ip2 \qquad (2)$$

$Ip1$ and $Ip2$ are pixel intensity of point P1 and P2. $Ip$ is the pixel intensity of P. $Ip$ linearly interpolates and fills pixels between P1 and P2.

Figure 4 illustrates these two equations. Figure 5 is an example using pixel intensity interpolation to obtain successive pixel intensity values between two regions.

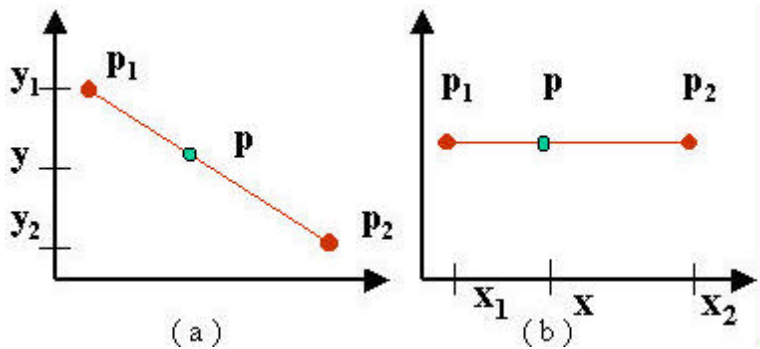

Figure 4: Illustration of equations. (a) for equation (1). (b) for equation (2).

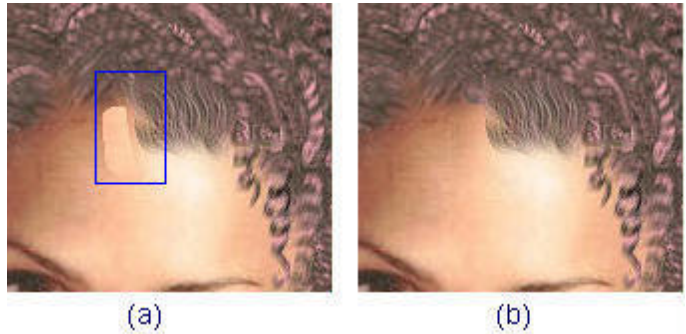

Figure 5: Pixel intensity interpolation. (a) In the region of blue square, the pixel change is not smooth. (b) The result after interpolation.

## 3. Result

Several synthetic results of the method are displayed in this section. Figure 6 and 7 show the change of the original picture's hairstyles to new hairstyles and the generation of facial hair. The results achieved the realism of images. Figure 6 (c) and Figure 7 (b) exhibit the photorealistic straight hair. Figure 6 (b) shows the generated curly hair. Figure 7 (c) shows the generated beard and mustache on a real person's face.

## 4. Conclusions and Future Work

In this paper, the method is presented, which can be used to fast and easily generate photorealistic hair. Although all examples in this paper are human hair, the method can be applied to other objects, such as fur, grass, etc. The future work is to develop a system for automatically designing and generating 2D photorealistic hair and to develop a method for generating photorealistic 3D-hair from a range of 2D photorealistic images.

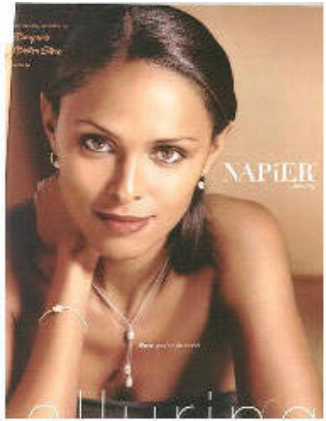
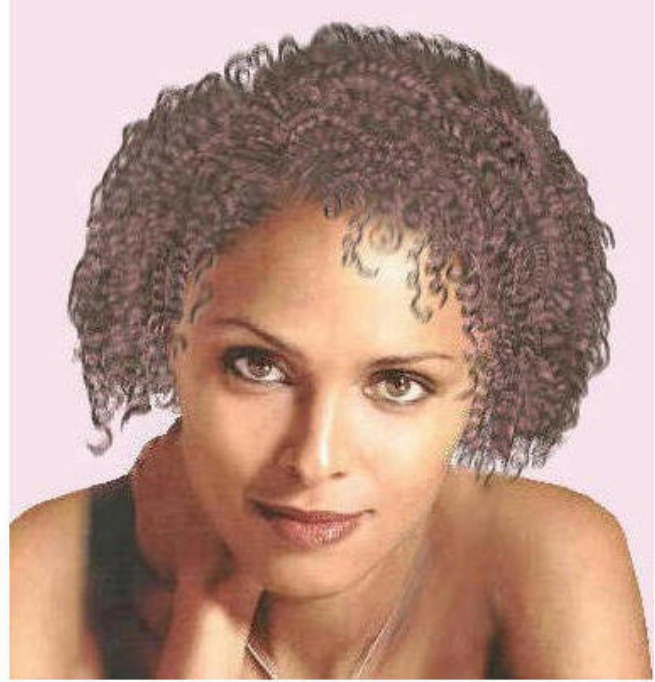
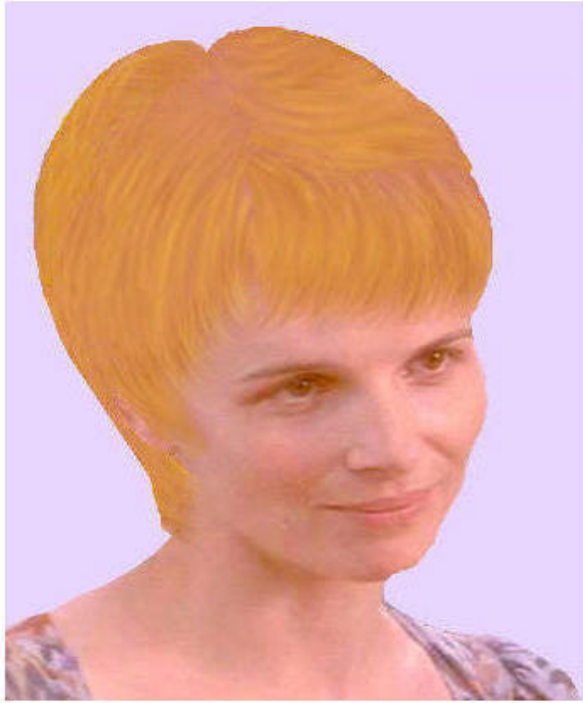
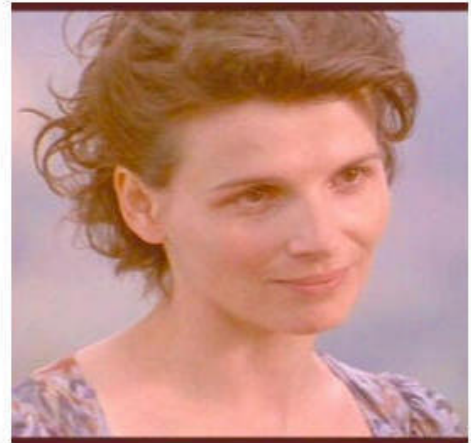

Figure 6: Generating straight and curly hairstyles. (a) Original image [4]. (b) Created curly hairstyle. (c) Created straight new hairstyle. (d) Original image [3]

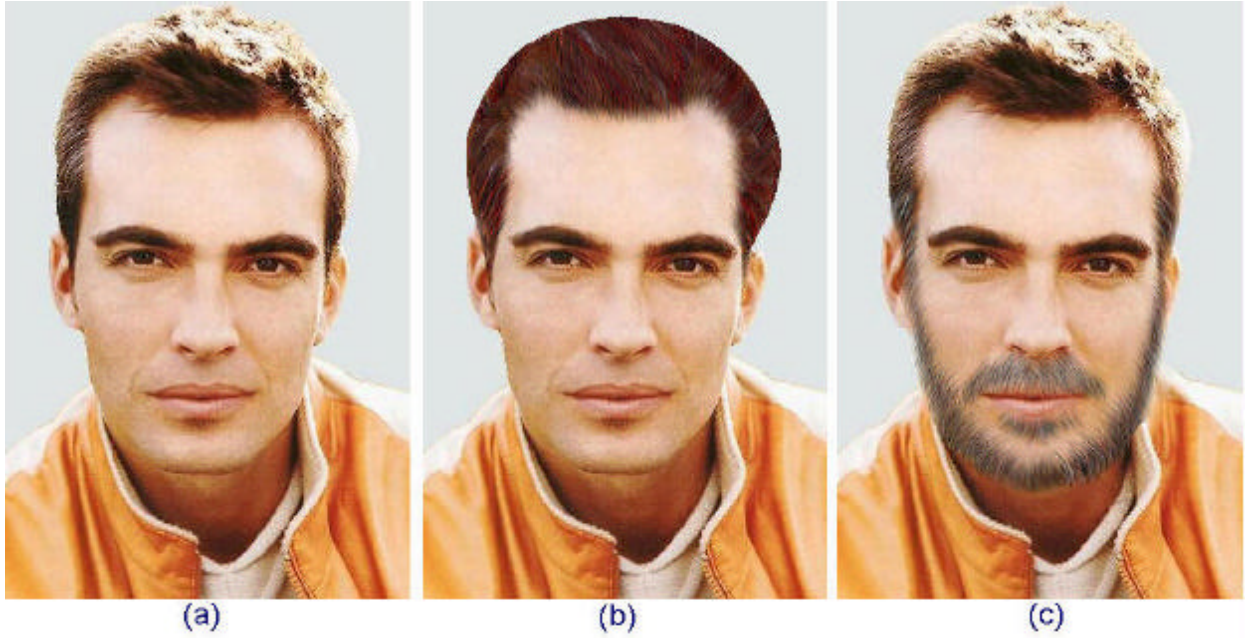

Figure 7: Hair and facial hair generation. (a) Original image [5]. (b) Created new hairstyle. (c) Created beard and mustache.